\documentclass[apj]{emulateapj} 
\usepackage{apjfonts} 
\usepackage{graphicx}
\def\alwaysmath#1{\ifmmode{#1}\else{$#1$}\fi}

\lefthead{Hasselquist et al.} 
\righthead{Chemical Abundances of the Sagittarius Dwarf Galaxy}
\begin{document} 
\DeclareGraphicsExtensions{.ps,.pdf,.png,.jpg,.eps}

\title{APOGEE Chemical Abundances of the Sagittarius Dwarf Galaxy} 
 
\author{Sten Hasselquist\altaffilmark{1}, 
Matthew Shetrone\altaffilmark{2}, 
Verne Smith\altaffilmark{3}, 
Jon Holtzman\altaffilmark{1}
Andrew McWilliam\altaffilmark{4}, 
J. G. Fern\'andez-Trincado\altaffilmark{5},
Timothy C. Beers\altaffilmark{6},
Steven R. Majewski\altaffilmark{7},   
David L. Nidever\altaffilmark{3}, 
Baitian Tang\altaffilmark{5},
Patricia B. Tissera\altaffilmark{8},
Emma Fern{\'a}ndez Alvar \altaffilmark{9},
Carlos Allende Prieto\altaffilmark{10,11},
Andres Almeida\altaffilmark{12},
Borja Anguiano\altaffilmark{7},
Giuseppina Battaglia\altaffilmark{10,11},
Leticia Carigi\altaffilmark{9},
Gloria Delgado Inglada \altaffilmark{9},
Peter Frinchaboy\altaffilmark{13}, 
D. A. Garc{\'{\i}}a-Hern{\'a}ndez\altaffilmark{10,11},
Doug Geisler \altaffilmark{5},
Dante Minniti\altaffilmark{14,15,16},
Vinicius M. Placco\altaffilmark{6}, 
Mathias Schultheis\altaffilmark{17},
Jennifer Sobeck\altaffilmark{7},
Sandro Villanova\altaffilmark{5}
}

\altaffiltext{1}{New Mexico State University, Las Cruces, NM 88003, USA (sten, holtz@nmsu.edu)}

\altaffiltext{2}{University of Texas at Austin, McDonald Observatory, Fort Davis, TX 79734, USA
(shetrone@astro.as.utexas.edu)}

\altaffiltext{3}{National Optical Astronomy Observatories, Tucson, AZ 85719, USA (vsmith, nidever@email.noao.edu)}

\altaffiltext{4}{The Observatories of the Carnegie Institute of Washington, 813 Santa Barbara Street, Pasadena, CA 91101, USA; andy@obs.carnegiescience.edu}

\altaffiltext{5}{Departamento de Astronom\'ia, Casilla 160-C, Universidad de Concepci\'on, Concepci\'on, Chile  (jfernandezt@astro-udec.cl, tangbaitian@gmail.com, dgeisler@astroudec.cl)}

\altaffiltext{6}{Department of Physics and JINA Center for the Evolution of the Elements,
University of Notre Dame, Notre Dame, IN 46556, USA  (tbeers, vplacco@nd.edu)}

\altaffiltext{7}{Dept. of Astronomy, University of Virginia, 
Charlottesville, VA 22904-4325, USA (srm4n, ba7t, jsobeck@virginia.edu)}

\altaffiltext{8}{Department of Physics, Universidad Andres Bello, 700 Fernandez Concha, Chile (tissera.patriciab@gmail.com)}

\altaffiltext{9}{Instituto de Astronom\'ia, Universidad Nacional Autónoma de M\'exico, Apdo. Postal 70264, Ciudad de M\'exico, 04510, Mexico (emma, carigi gdelgado@astro.unam.mx)}

\altaffiltext{10}{Instituto de Astrof{\'{\i}}sica de Canarias, E-38205 La Laguna, Tenerife, Spain (callende, agarcia@iac.es)}

\altaffiltext{11}{Departamento de Astrof{\'{\i}}sica, Universidad de La Laguna (ULL), E-38206 La Laguna, Tenerife, Spain (callende, agarcia@iac.es, battagliagemr@gmail.com)}

\altaffiltext{12}{Departamento de F{\'{\i}}sica, Facultad de Ciencias, Universidad de La Serena, Cisternas 1200, La Serena, Chile (aalmeida@userena.cl)}

\altaffiltext{13}{Texas Christian University, Fort Worth, TX 76129, USA (p.frinchaboy@tcu.edu)} 

 \altaffiltext{14}{Departamento de Fisica, Facultad de Ciencias Exactas, Universidad Andres Bello, Av. Fernandez Concha 700, Las Condes, Santiago, Chile. (vvvdante@gmail.com)}
 
\altaffiltext{15}{Instituto Milenio de Astrofisica, Santiago, Chile. (vvvdante@gmail.com)}

\altaffiltext{16}{Vatican Observatory, V00120 Vatican City State, Italy. (vvvdante@gmail.com)}

\altaffiltext{17}{Laboratoire Lagrange, Universit\'e C\^ote d'Azur, Observatoire de la C\^{o}te d'Azur, CNRS, Bd de l'Observatoire, 06304, Nice, France (mathias.schultheis@oca.eu)}

\begin{abstract} 
The Apache Point Observatory Galactic Evolution Experiment (APOGEE) provides the opportunity to measure elemental abundances for C, N, O, Na, Mg, Al, Si, P, K, Ca, V, Cr, Mn, Fe, Co,  and Ni in vast numbers of stars. We analyze the chemical abundance patterns of these elements for 158 red giant stars belonging to the Sagittarius dwarf galaxy (Sgr). This is the largest sample of Sgr stars with detailed chemical abundances and the first time C, N, P, K, V, Cr, Co, and Ni have been studied at high-resolution in this galaxy. We find that the Sgr stars with [Fe/H] $\gtrsim$ -0.8 are deficient in all elemental abundance ratios (expressed as [X/Fe]) relative to the Milky Way, suggesting that Sgr stars observed today were formed from gas that was less enriched by Type II SNe than stars formed in the Milky Way. By examining the relative deficiencies of the hydrostatic (O, Na, Mg, and Al) and explosive (Si, P,  K, and Mn) elements, our analysis supports the argument that previous generations of Sgr stars were formed with a top-light IMF, one lacking the most massive stars that would normally pollute the ISM with the hydrostatic elements. We use a simple chemical evolution model, flexCE to further backup our claim and conclude that recent stellar generations of Fornax and the LMC could also have formed according to a top-light IMF.
\end{abstract}

 
\section{Introduction}

The Sagittarius dwarf galaxy (Sgr) is a nearby, massive dwarf  galaxy that is currently merging with the Milky Way (MW). Discovered by \citet{Ibata1994}, Sgr is one of the MW's most massive satellites and as a consequence of the merger, massive tidal tails belonging to Sgr can be found in the Galactic halo (e.g., \citealt{Ibata2001,Majewski2003,Belokurov2006}). The main body of Sgr contains its own system of globular clusters, and several globular clusters in the MW's halo have been associated with the Sgr stream \citep{Law&Majewski2010b}. Understanding the complex Sgr system can help answer questions about the formation of our Galactic halo, first suggested by \citet{Searle&Zinn1978} to have formed through the accretion of sub-galactic ``fragments''. Whether these ``fragments'' were dwarf galaxies, such as Sgr, or other stellar systems, is still an open question (see e.g., \citealt{Venn2004,Geisler2007}).

Because it is relatively nearby (d $\sim$ 27-28 kpc, \citealt{Siegel2011,Hamanowicz2016}), the stars in the core of Sgr are more readily accessible to high-resolution spectroscopy from the ground than for other MW dSph galaxies. There have been a number of spectroscopic studies characterizing the dwarf's chemical properties (\citealt{Bonifacio2000,Smecker-Hane&McWilliam2002,Bonifacio2004,Chou2007,Sbordone2007,Carretta2010,Chou2010,Majewski2013,McWilliam2013,Mucciarelli2017}). These studies all show that Sgr is the most metal rich of all MW dSph galaxies, with a mean [Fe/H] $\sim$ -0.5, but also posseses a large spread in metallicity (-1.4 < [Fe/H] < 0.1). \citet{Siegel2007} analyzed Hubble Space Telescope CMDs to determine that Sgr harbors up to five distinct populations spanning a wide range of age as well (2 --- 13 Gyr). A range of populations are known to exist in other MW dSphs such as Sculptor (e.g., \citealt{Tolstoy2004}) and Fornax (e.g., \citealt{Battaglia2006}). Even though the MW dSphs do not show any signs of star formation today, it is clear that they have had a complex star formation history (e.g., \citealt{Mateo1998,Dolphin2002}).

The most comprehensive, detailed chemical abundance study of Sgr to date is that of \citet{McWilliam2013}. They derived detailed chemical abundances for three stars, and combined their sample with those of \citet{Smecker-Hane&McWilliam2002}, \citet{Bonifacio2004}, and \citet{Sbordone2007} to produce a sample of $\sim$ 30 Sgr members with detailed chemical abundances. \citet{McWilliam2013} argued that the chemical abundance ratios of Sgr (expressed as [X/Fe]), which include deficiencies in most elements relative to the MW (aside from those created by the $s$-process), indicate that the most recent generations of Sgr stars formed according to a top-light IMF, one lacking the more massive molecular clouds needed to form the most massive stars (e.g., \citealt{Oey2011,Kroupa2013}). Evidence for this stems from their analysis which showed that Sgr is not equally deficient in all chemical elements (including the $\alpha$-elements: O, Mg, Si, S, Ca, and Ti), and in fact appears to be more deficient in elements produced in hydrostatic burning phases in massive stars (e.g., O, Mg, and Al) than elements produced in explosive nucleosynthesis preceding Type II SNe (e.g., Si, Ca, and Ti). The nucleosynthetic yields of the hydrostatic elements are more mass-dependent than the yields of the explosive elements \citep{Woosley&Weaver1995}.

Studies of other classical dSph galaxies find that the more metal-rich stars have $\alpha$-element abundance ratios that fall below the MW trend (\citealt{Shetrone2001,Shetrone2003,Venn2004,Geisler2005}). This has been attributed to low-efficiency star formation that enriches the dSph ISM to a much lower metallicity than the MW before the initiation of Type Ia SNe. However, to have $\alpha$-element abundances that are more deficient than MW stars after the turn on of Type Ia SNe requires that the low-efficiency star formation also causes lowered $\alpha$-enhancement for the metal-poor stars. This can happen if the dwarf galaxies are deficient in ejecta from massive Type II SNe relative to the MW. \citet{Lemasle2014} concluded that the low [$\alpha$/Fe] and enhanced [Eu/Mg] abundance ratios of the younger Fornax populations indicate that these stars formed from gas lacking in ejecta from massive Type II SNe. Similar results for the Large Magellanic Cloud have been found by \citet{VanderSwaelmen2013}; they concluded that massive stars were less important for the chemical enrichment of the LMC than that of the MW. The Small Magellanic Cloud, which has a similar mass to the pre-stripped Sgr \citep{Law&Majewski2010} but is still forming stars, shows signs of a top-light IMF, but no obvious cutoff of massive stars in an analysis of a sample of OB stars from \citet{Lamb2012}.

Large-scale surveys of dSphs are required to achieve a better understanding of the chemical evolution of these systems. Unfortunately, these galaxies are sufficiently faint that high-resolution spectroscopic surveys capable of deriving detailed chemical abundances require large amounts of large-aperture telescope time; thus large samples (> 100 stars) do not yet exist. However, the Apache Point Observatory Galactic Evolution Experiment (APOGEE) has already shown that precise radial velocities and metallicities can be obtained for Sgr members \citep{Majewski2013}. APOGEE is a survey that employs a high resolution ($R$ $\sim$ 22,500), $H$-band spectrograph to obtain spectra for hundreds of thousands of red giants across the MW \citep{Majewski2017}.  The APOGEE survey in SDSS-III \citep{Eisenstein2011} observed five fiber plug-plates in the direction of Sgr and over 300 Sgr members were confirmed by \citet{Majewski2013}. APOGEE can derive accurate stellar parameters as well as reliable chemical abundances for 18 elements from carbon up through the iron peak.

In this paper, we analyze the APOGEE observations for 158 confirmed Sgr red giants in the core regions (within 5 deg) of Sgr. Observations and data reduction are described in \S \ref{obs}. We present the detailed chemical abundance results in \S \ref{res}, where we find that Sgr stars more metal rich than [Fe/H] = -0.8 exhibit deficiencies in chemical-abundance ratios (expressed as [X/Fe]) compared to similar metallicity stars in the MW. We employ a chemical-evolution model (flexCE; \citealt{Andrews2017}) to analyze how changing parameters in the star-formation history of Sgr affects the chemical abundance patterns. The model and results are presented and analyzed in \S \ref{flex_ce}. We find that the APOGEE abundance results for Sgr stars with [Fe/H] > -0.8 are consistent  with the top-light IMF scenario hypothesized in \citet{McWilliam2013} and that it is unlikely that stellar winds and/or low star-formation efficiency can explain the measured abundance patterns.

\section{Observations, Data Reduction, Analysis}
\label{obs}

APOGEE was part of Sloan Digital Sky Survey III \citep{Eisenstein2011}, and observed 146,000 stars in the Milky Way galaxy \citep{Majewski2017}. The APOGEE instrument is a high-resolution ($R\sim$ 22,500) near-infrared (1.51-1.70 $\mu$m) spectrograph described in detail in Wilson et al. (in prep). For the main survey, the instrument was connected to the Sloan 2.5m telescope \citep{Gunn2006}. The APOGEE data are reduced through methods described by \citet{Nidever2015} and stellar parameters are extracted using the APOGEE Stellar Parameters and Chemical Abundances Pipeline (ASPCAP, \citealt{Garcia2016}). ASPCAP interpolates in a grid of synthetic spectra \citep{Zamora2015} to find the best fit (through $\chi^{2}$ minimization) to the observed spectrum by varying effective temperature, surface gravity, metallicity, carbon abundance, nitrogen abundance, and $\alpha$-element abundance. Microturbulence is determined through an empirically found relation with surface gravity (log(g)), as derived for a subset of APOGEE stars used for calibration. In this analysis we use results from the 13th data release of SDSS (DR13, \citealt{Albareti2017} and \citealt{Holtzman:inprep-c}). 

DR13 also provides measurements of individual chemical abundances for C, N, O, Na, Mg, Al, Si, P, S, K, Ca, Ti, V, Cr, Mn, Fe, Co,  and Ni.  In this analysis, we generally report values relative to iron ([X/Fe]) rather than absolute abundances because the relative abundances are more useful when exploring the relative contributions of SNe and AGB stars to the gas that formed the most recent generation of stars. The uncertainties of the individual chemical abundance measurements from DR13 are estimated from the scatter of chemical abundances within open clusters in bins of S/N and effective temperature (see \citealt{Holtzman:inprep-c} or the DR13 webpages\footnote{http://www.sdss.org/dr13/irspec/abundances/} for a complete description). 

The values we adopt come from DR13 (\citealt{Holtzman:inprep-c}), except for K. At the Sgr velocity, one of two K features in the APOGEE spectral region (15167.081 \AA ) falls beneath a bright sky line that is not properly subtracted in all cases. This causes added uncertainty to the ASPCAP K abundances. Therefore, we rederived abundances using just the line at 15172.572 \AA $ $ using Turbospectrum synthesis \citep{Alvarez1998}, adopting the stellar parameters from ASPCAP.

\begin{figure*}
\plottwo{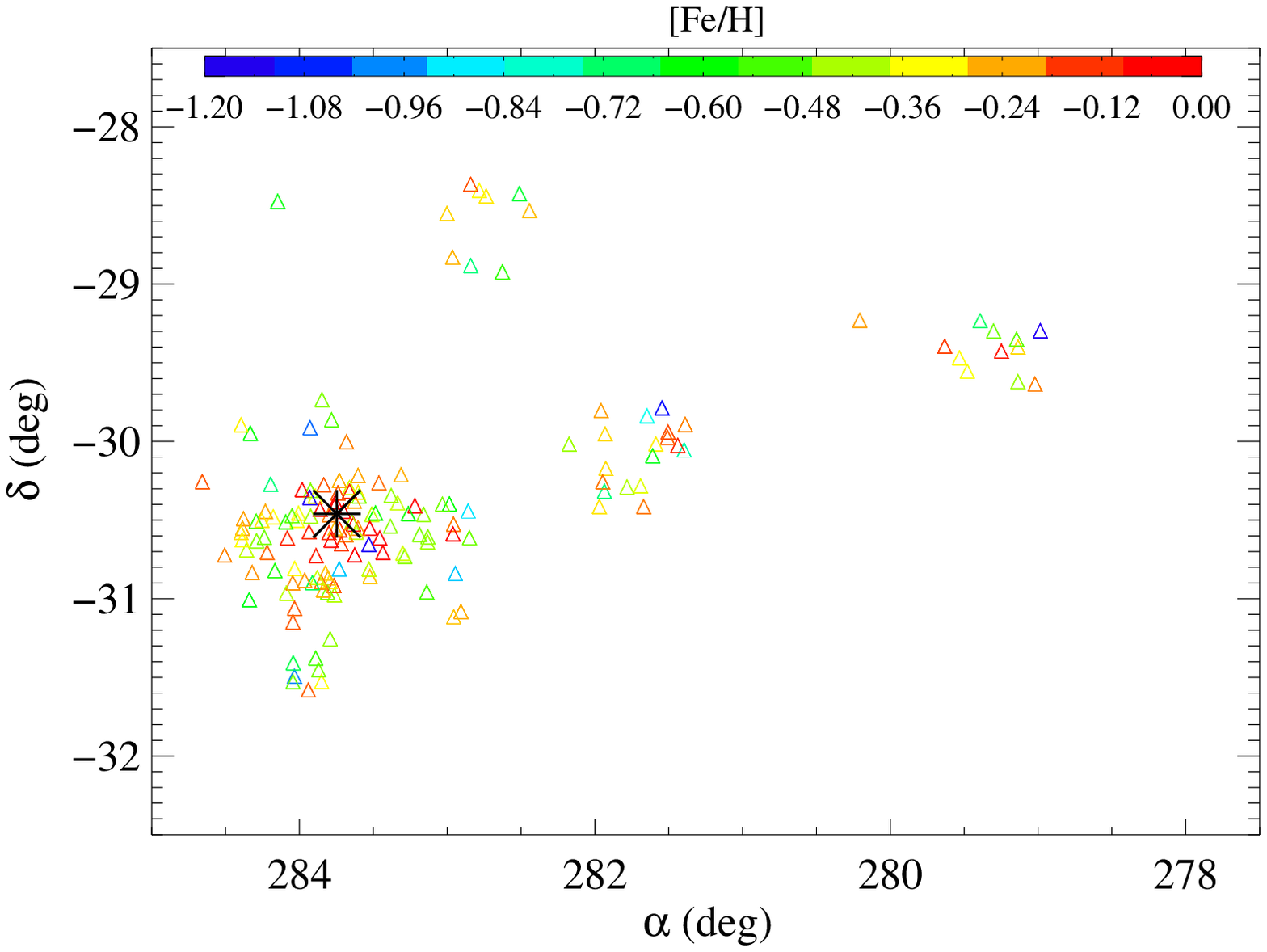}{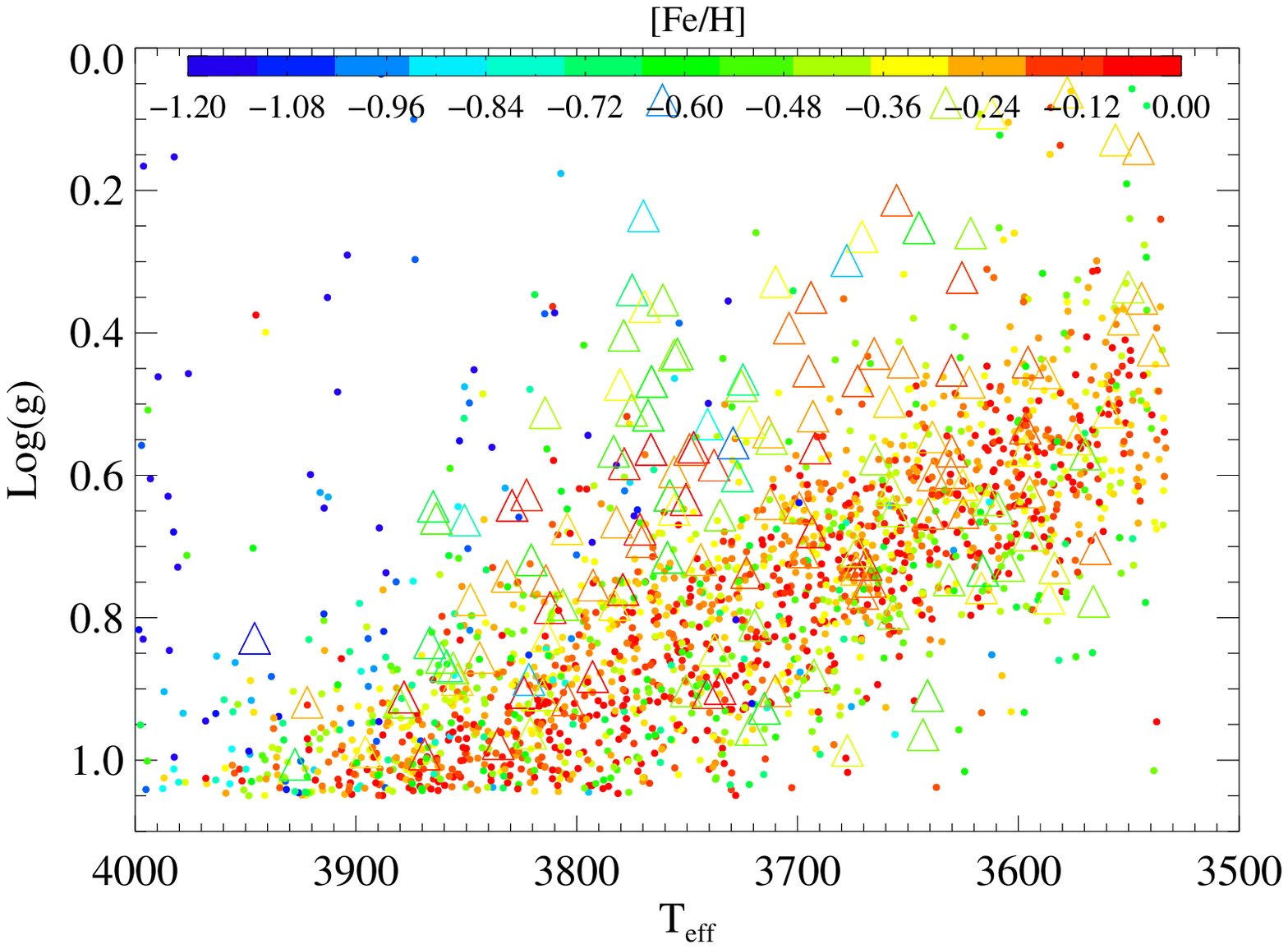}
\caption{Left Panel: sky distribution of Sgr stars studied in this work. The black asterisk indicates the center of Sgr as given by \citet{Majewski2003}}. Points are color-coded by [Fe/H], as indicated in the color bar. Right panel: HR diagram of the Sgr stars (colored open triangles) and the MW comparison samples (colored filled circles). Points are color-coded by [Fe/H], as indicated in the color bar.
\label{cmd_dist}
\end{figure*}

Sgr members were determined based on an iterative RV selection method along with a CMD selection to isolate the red giant branch of the Sgr core, described in detail in \citet{Majewski2013}. Four more Sgr fiber plug-plates have since been observed resulting in a Sgr sample of 158 stars with spectra having S/N > 80. The sky distribution of Sgr stars is shown in the left panel of Figure \ref{cmd_dist}. The majority of the Sgr sample lies within the inner 1 degree of the center of Sgr, but we do have a small handful of stars $\sim$ 5 degrees away from the center, along the major axis. These stars are outside the core radius of 224' given by \citet{Majewski2003}.

In our abundance analysis, we select a MW sample that we use to compare Sgr to. This MW sample is selected to cover the same parameter space as Sgr to minimize the influence of possible systematic effects on the comparison between the MW and Sgr chemical-abundance patterns:

\begin{itemize}
\item 3500 < T\raisebox{-.4ex}{\scriptsize eff} < 4000 K
\item log(g) < 1.05
\item -1.2 < [Fe/H] < 0.0
\item S/N > 80
\item No ``ASPCAPBAD'' flag set\footnote{This flag, described in detail in \citet{Holtzman2015}, indicates whether a star falls near edges of the synthetic spectra grids, or has a low $S/N$ spectrum.}
\end{itemize}

Figure \ref{cmd_dist}, right panel shows the HR diagram for the Sgr sample and the MW sample. While it's clear that Sgr and MW cover similar values of log(g) and T\raisebox{-.4ex}{\scriptsize eff}, there is  a systematic offset between the two giant branches. This could be due to the fact that Sgr is $\alpha$-element poor relative to the MW stars (shown in Figure \ref{intro_plot}), which pushes the giant branch to hotter temperatures on the HR diagram.

The MW sample is divided spatially into disk, bulge, and high-latitude (Table \ref{MW_samp}). The high-latitude sample was designed to select stars residing in the MW stellar halo, but contains some thin/thick-disk contamination. In the following sections, the points in the figures are color-coded as indicated by Table \ref{MW_samp} unless otherwise noted.  

\begin{table}[htp]
 \caption{MW Comparison Samples}
\begin{center}
\begin{tabular}{ l c c r }
\hline
 & \textbf{Disk} & \textbf{Bulge} & \textbf{High-latitude}\\
 
\hline
\textbf{b\raisebox{-.4ex}{\scriptsize cut}} & |$b$| < $4^{\circ}$ & |$b$| < $4^{\circ}$ & |$b$| > $40^{\circ}$ \\
\textbf{l\raisebox{-.4ex}{\scriptsize cut}}  & |$l$| > $10^{\circ}$ & $4^{\circ}$ < $l$ < $-4^{\circ}$ & |$l$| > $20^{\circ}$\\
\textbf{N\raisebox{-.4ex}{\scriptsize stars}} & 1384 & 431 & 149\\
\textbf{Plot Color} & black & red & green\\
\tableline
\end{tabular}
\\
\end{center}
\label{MW_samp}
\end{table}




\section{Results}
\label{res}

\begin{figure}[h]
\epsscale{1.0}
\plotone{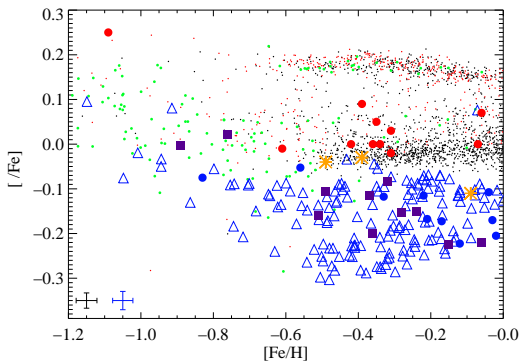} 
\caption{The [$\alpha$/Fe] vs. [Fe/H] abundance plane for the APOGEE MW disk (black dots), APOGEE MW bulge (red dots), APOGEE MW halo (green dots), APOGEE Sgr sample (blue open triangles), \citet{McWilliam2013} (orange stars), \citet{Smecker-Hane&McWilliam2002} (red, filled circles), \citet{Bonifacio2004} (blue, filled circles), and \citet{Sbordone2007} (magenta squares).}
\label{intro_plot}
\end{figure}

We first present the results for [Fe/H] and $\alpha$-elements ([$\alpha$/Fe]) derived by APOGEE (Fig \ref{intro_plot}). Several studies (e.g., \citealt{Bonifacio2004,McWilliam2013}, and references therein) have shown that the more metal-rich Sgr stars are deficient in [$\alpha$/Fe] relative to MW stars. In Figure \ref{intro_plot} we include the sample of Sgr stars analyzed by \citet{McWilliam2013}, and find that the APOGEE [$\alpha$/Fe] abundance patterns are similar to those shown by the \citet{McWilliam2013} total sample, with some stars in the \citet{Smecker-Hane&McWilliam2002} sample (red circles) falling at slightly higher $\alpha$-element abundances than the APOGEE Sgr metal-rich stars. \citet{McWilliam2013} found that the \citet{Bonifacio2004} and \citet{Sbordone2007} samples seemed to be a little too deficient when compared to their three stars and the stars from \citet{Smecker-Hane&McWilliam2002} whereas the APOGEE stars actually agree well with the lower \citet{Bonifacio2004} and \citet{Sbordone2007} stars. While the \citet{McWilliam2013} stars are in the center of Sgr and the samples of \citet{Bonifacio2004} and \citet{Sbordone2007} are outside the core, in \S \ref{comp} we explain that the differences in [$\alpha$/Fe] between samples are not likely to be a result of sampling different areas of the Sgr core. We find Sgr to be deficient in the $\alpha$-elements across the space sampled by APOGEE. 

It is possible that the APOGEE data suffer from systematic zero-point offsets, as we are working in the lower temperature regime where the ASPCAP abundances are more uncertain than for stars with temperatures above 4000 K, which could explain the discrepancy. For example, The mean [$\alpha$/Fe] abundance for MW disk stars with [Fe/H] = 0.0 observed by APOGEE is slightly sub-solar ([$\alpha$/Fe $\sim$ -0.02) whereas \citet{Edvardsson1993} finds a slightly super-solar mean [$\alpha$/Fe] ([$\alpha$/Fe $\sim$ +0.03). Comparing the 3 \citet{McWilliam2013} stars to a MW disk at <[$\alpha$/Fe]> = +0.03 would indicate that they are deficient by about 0.05 dex. The discrepancy between the \citet{Smecker-Hane&McWilliam2002} sample and the other Sgr samples can not be explained in this way. However, in this study, because we are comparing stars with very similar stellar parameters within the APOGEE sample, any systematic uncertainties should have limited effect on the relative abundance distributions between the MW and Sgr.

The deficiencies in the $\alpha$-elements observed in the APOGEE Sgr sample suggest that the most recent generations of Sgr stars were formed from an ISM lacking ejecta from Type II SNe. \citet{McWilliam2013} points to an IMF lacking in the most massive stars (top-light IMF), either via a steeper IMF slope (e.g., \citealt{Oey2011}) or an upper mass cutoff above which stars could not form (e.g., \citealt{Weidner&Kroupa2005}), as the explanation of these deficiencies. This scenario can explain the relative deficiencies between the hydrostatic elements, the yields of which are mass dependent, and the explosive elements (see \citealt{Woosley&Weaver1995} and \citealt{Nomoto2006}). In the following subsections, we analyze the individual chemical-abundance patterns for each group of elements that APOGEE provides, which are divided here as CNO, $\alpha$-elements, odd-$Z$ elements, and Fe-peak elements, and discuss whether the detailed chemical-abundance patterns of Sgr are consistent with the top-light IMF hypothesis of \citet{McWilliam2013}.

\subsection{C, N, and O}
\label{cno}

Carbon, nitrogen, and oxygen are some of the most abundant elements in the Universe after H and He, but their study in the context of galaxy chemical evolution has been hampered due to observational difficulties in the optical and the complication of abundance variations arising from stellar evolution rather than primordial variation. Fortunately, the $H$-band (1.5 - 1.7 $\mu$m) is rich with molecular features (CO, CN, and OH) from which APOGEE can derive highly precise (random uncertainties of $\sim$ 0.03 dex) C, N, and O abundances. C and N are both elements that have not been extensively studied in Sgr at high spectra resolution. \citet{McDonald2012} analyzed CN band strengths of $\sim$ 900 Sgr members using low-resolution spectroscopy to measure the fraction of carbon stars in Sgr, and found evidence for such stars acting as important producers of the carbon in the ISM of Sgr, but the authors do not comment on the C abundance pattern of Sgr relative to the MW.  Nitrogen has not been previously studied in Sgr.

\begin{figure*}[h]
\epsscale{1.}
\plotone{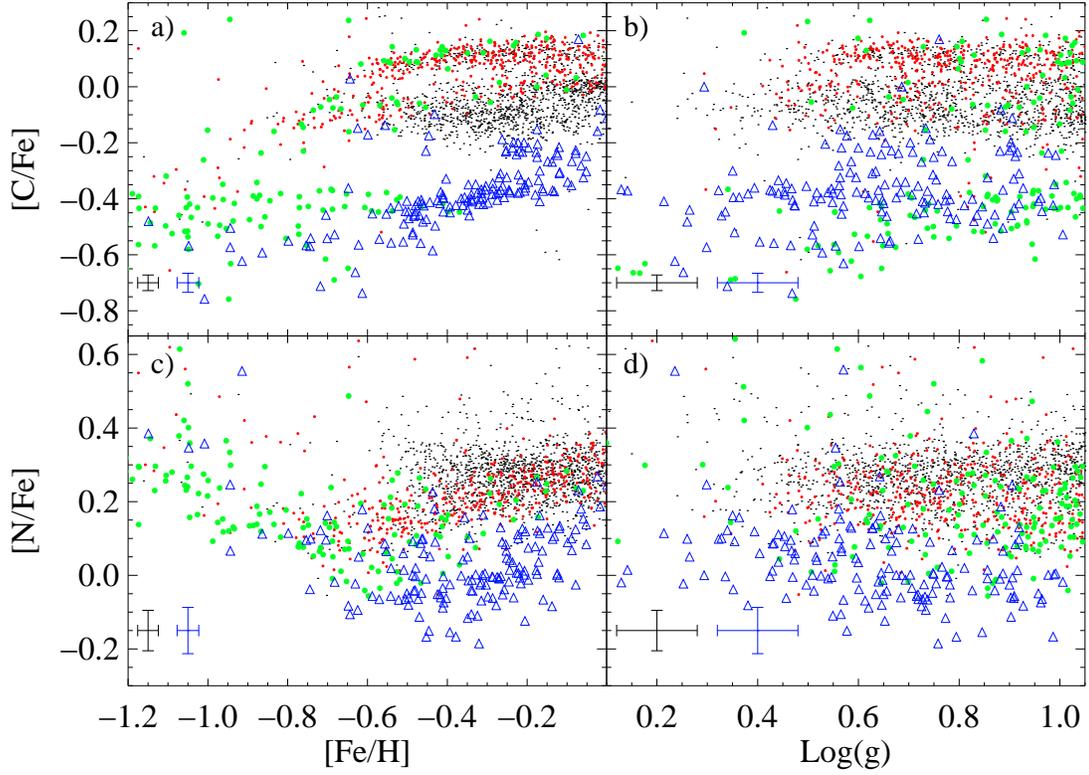} 
\caption{Plots of the abundances of C, N, and O. Sgr members are plotted as blue open triangles. The MW disk (black dots), MW bulge (red dots), and MW high-latitude (green filled circles) samples are also plotted.  Median 1-$\sigma$ error bars for the MW sample (black) and Sgr sample (blue) are indicated in the lower left each panel.}
\label{cno_plot}
\end{figure*}

\begin{figure*}[h]
\epsscale{1.}
\plotone{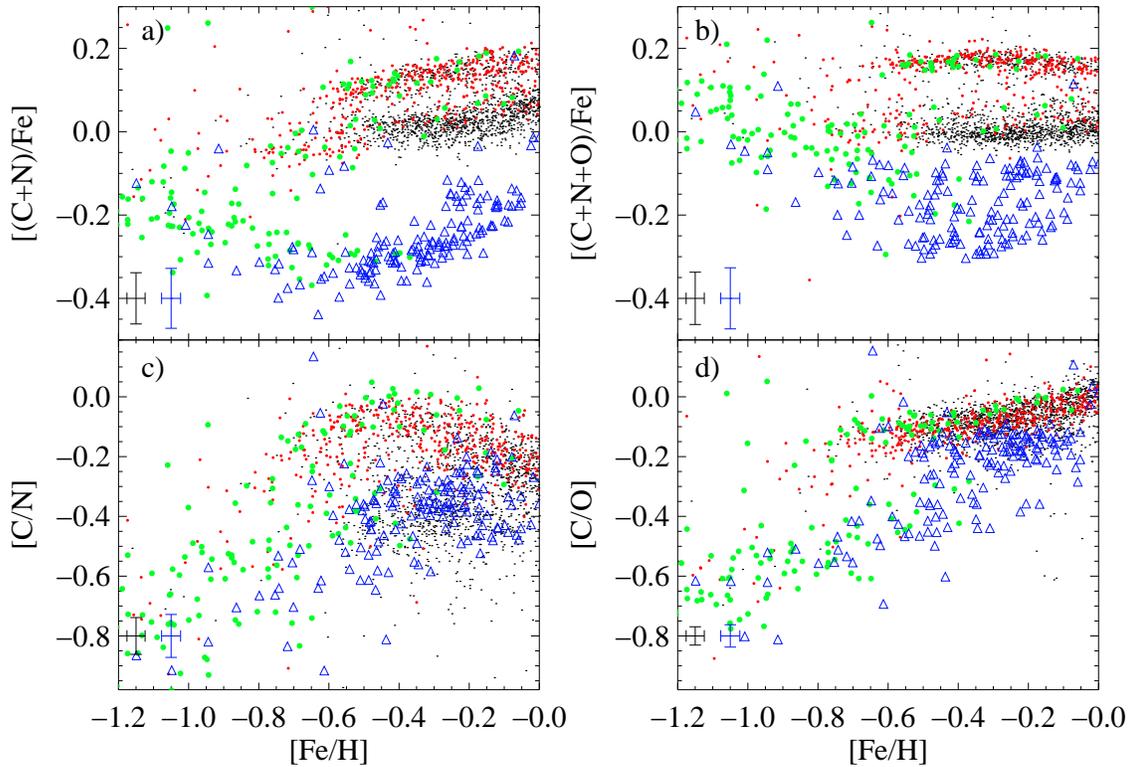} 
\caption{Plots of the abundances of C+N, C+N+O, C/N, and C/O. Points are color-coded the same way as in Figure \ref{cno_plot}.}
\label{cno_plot_b}
\end{figure*}

Figure \ref{cno_plot}, panels a and c, show the CN abundance patterns for Sgr. Sgr is deficient relative to the MW in [C/Fe] by $\sim$ 0.4 dex for stars with [Fe/H] > -1.0 and in [N/Fe] by $\sim$ 0.2 dex for stars with [Fe/H] > -0.6. Sgr stars with [Fe/H] < -0.8 are deficient relative to the high-latitude sample by $\sim$ 0.1 dex for [C/Fe] and not deficient in [N/Fe]. To show that these differences are not a result from observing red giants that have gone through different amounts of CN processing, we also plot [C/Fe] and [N/Fe] vs. log(g) in the second column of Figure \ref{cno_plot}. We find that for any given log(g), the Sgr [C/Fe] and [N/Fe] abundance ratios for the more metal-rich Sgr stars are lower than the MW. Because CN processing generally enhances nitrogen in the atmospheres of red giants and dilutes carbon \citep{Iben1964}, [(C+N)/Fe] is a more robust indicator of the primordial C and N abundance. In the third row of Figure \ref{cno_plot_b}, we show that Sgr is deficient in [(C+N)/Fe] and [(C+N+O)/Fe], which indicates that Sgr stars formed from gas that was deficient in these light elements.

The [(C+N)/Fe] abundances exhibit a correlation with [Fe/H] starting at [Fe/H] $\sim$ -0.6 and an anti-correlation at [Fe/H] < -0.8 where the Sgr sample joins with the MW high-latitude sample. \citet{McWilliam2013} found an increase of [La/Fe] in Sgr stars beginning at [Fe/H] $\sim$ -0.6. C, N, and the $s$ process elements are produced in AGB stars (see yields from \citealt{Karakas2014} and PNe measurements from e.g. \citealt{Sharpee2007,Garcia-Rojas2012,Garcia-Rojas2015}). O has also been found to be produced by AGB stars, but in lower quantities than the previously mentioned elements (e.g., \citealt{Delgado-Inglada2015}). The [(C+N)/Fe] and [La/Fe] abundance patterns suggest that AGB stars were more important polluters of the ISM relative to SNe than in the MW for stars with [Fe/H] > -0.6. The anticorrelation at [Fe/H] < -0.8 is due to Type Ia SNe lowering the [(C+N)/Fe] abundance ratio before reaching a metallicity at which AGB stars more effectively produce C and N, as is seen in other dwarf galaxies (see e.g., \citealt{Kirby2015}).

It has been shown that the [C/N] abundance ratio can be used as a mass indicator for red giant stars, from which the age can be inferred (e.g., \citealt{Salaris2015,Martig2016}). While the temperatures and surface gravities of the APOGEE MW and Sgr samples are outside the calibration ranges for the [C/N]-age relations from the literature, we show in panel c of Figure \ref{cno_plot_b} that Sgr stars with [Fe/H] > -0.6 exhibit similar [C/N] abundance patterns as the MW thin-disk stars. This suggests that the more metal-rich Sgr stars have similar ages to the MW thin disk. This is consistent with \citet{Siegel2007}, whose work suggests that the stars in our Sgr sample have ages as young as 2 Gyr and as old as 8 Gyr.

Low [(C+N+O)/Fe] is consistent with the top-light IMF scenario because the yields of these elements are mass-dependent in Type II SNe. C is more mass-dependent than O, which explains the [C/O] abundance pattern shown in the lower right of Figure \ref{cno_plot_b}. Stars formed at [Fe/H] $\sim$ -0.8 have lower [C/O] because they formed with less Type II SNe ejecta from massive stars. As AGB enrichment becomes an important contributor to the chemical enrichment of Sgr, the [C/O] ratio increases to just below the MW trend at [Fe/H] $\sim$ -0.1.

\subsection{$\alpha$-Elements}
\label{alpha}

The $\alpha$-elements are the even-$Z$ elements from O to Ti, and are primarily synthesized in massive stars (M $> $ 8 $M_{\odot}$).  The lighter $\alpha$-elements, O and Mg, are synthesized via hydrostatic C and Ne burning in massive stars and are released to the ISM through Type II SNe explosions. These reactions typically occur in the outer burning shells of massive stars, making their yields dependent on the mass of the progenitors (e.g., \citealt{Woosley&Weaver1995} and \citealt{Nomoto2006}). Si, S, Ca, and Ti are synthesized via explosive O and Si burning that occurs in the core region of the massive star. Because the amount of mass in the core of massive stars only weakly depends on progenitor mass, the yields of Si, S, Ca, and Ti are less dependent on progenitor mass.

\begin{figure*}[h]
\epsscale{1.}
\plotone{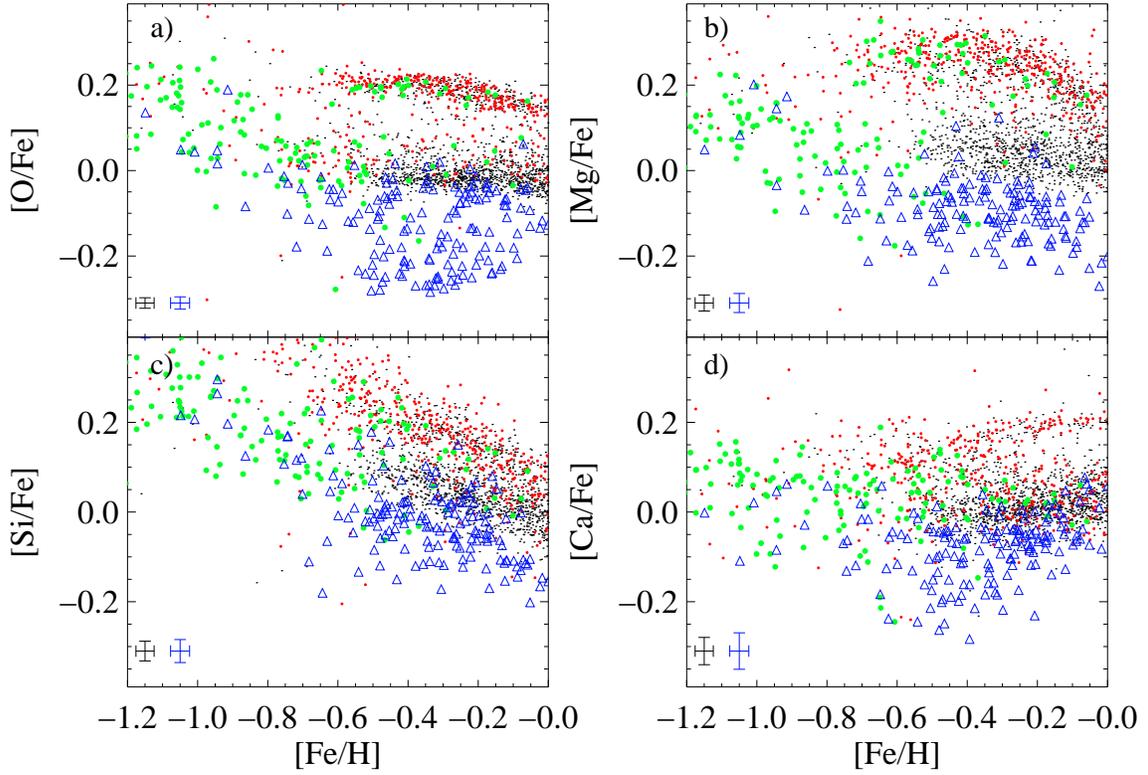} 
\caption{Plots of the $\alpha$ elements. Points are colored the same as in Figure \ref{cno_plot}. Median 1-$\sigma$ error bars are plotted in the lower left of each plot.}
\label{alphas_plot}
\end{figure*}

The $\alpha$-element abundance patterns are plotted in Figure \ref{alphas_plot}. Sgr is clearly deficient (by about 0.1 dex from the thin-disk trend) in both [O/Fe] and [Mg/Fe] for stars with [Fe/H] > -0.8. Sgr is also mildly deficient in [Si/Fe] --- by $\sim$ 0.05 dex. The larger deficiencies in [O/Fe] and [Mg/Fe], relative to the deficiency in [Si/Fe], is expected in the top-light IMF scenario. If the lack of Type II SNe ejecta were simply due to galactic winds or low star-formation efficiency, Sgr should have a similar magnitude of deficiencies across all $\alpha$-elements. The more metal-poor Sgr stars appear to join with the high-latitude MW stars in all $\alpha$ elements.

Similar $\alpha$-element deficiencies have been observed in other dwarf galaxies (e.g., Draco and Ursa Minor, \citealt{Shetrone2003},\citealt{Tolstoy2003}, and Fornax \citealt{Lemasle2014}). Such deficiencies have typically been attributed to the time-delay scenario of \citet{Tinsley1979} where the star formation rate is much lower in these galaxies than for the MW, such that fewer Type II SNe go off before the onset of Type Ia. However, this would only produce variance in the deficiencies of the $\alpha$-elements if the heavier $\alpha$-elements were produced in appreciable quantities in Type Ia SNe. While Type Ia SNe yields indicate that Si and Ca are produced by these objects (e.g., \citealt{Iwamoto1999} and \citealt{Maeda2010}), it is not clear that the yields are sufficient to produce the observed chemical-abundance patterns in Fornax and Sgr. However, $r$-process Eu has been studied in both Sgr and Fornax by \citet{McWilliam2013} and \citet{Lemasle2014}, respectively. Both of these galaxies show enhancements in [Eu/Mg] and [Eu/O] relative to the MW. The elements Eu, O, and Mg are not produced by the $r$-process in Type Ia SNe, so the fact that Sgr and Fornax are enhanced suggests that the $\alpha$-elements are lowered due to a mass-dependent paucity of Type II SNe ejecta, and not simply an excess of Type Ia ejecta.

\subsection{Odd-Z Elements}
\label{oddz_sec}

The odd-$Z$ elements, Na, Al, P, and K, are produced in various nuclear burning phases during the lifetime of a massive star. Na and Al are synthesized in the hydrostatic carbon and neon-burning stages of massive stars, whereas P and K are synthesized in the explosive burning stages. This implies that the yields of Na and Al are more mass-dependent than P and K. The yields of the odd-Z elements are also metallicity-dependent because the production of these elements occurs via reactions that rely on the presence of metals (e.g., \raisebox{+.4ex}{\scriptsize 14}N) in the burning layer that create a flux of neutrons. Na and Al can also be synthesized in AGB stars. AGB yields from \citet{Karakas2010}, incorporated into a simple chemical-evolution model of the MW in \citet{Andrews2017}, suggest that AGB stars become non-negligible contributors to the total Na and Al abundance in the MW for [Fe/H] > -0.2. 

\begin{figure*}[h]
\epsscale{1.}
\plotone{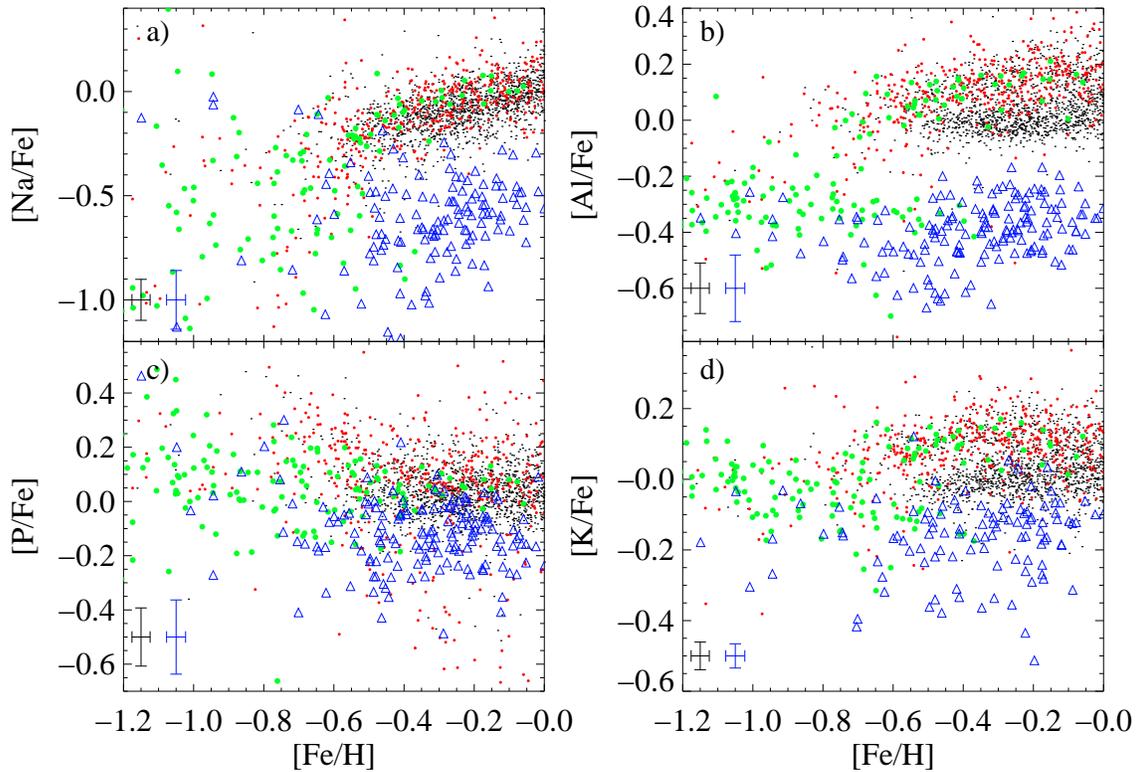} 
\caption{Plots of the odd-$Z$ element abundance patterns for Sgr and the MW. Points are colored the same as in Figure \ref{cno_plot}.}
\label{odd_z_plot}
\end{figure*}

The abundance patterns for the odd-Z elements in Sgr are shown in Figure \ref{odd_z_plot}. Sgr is deficient in [Na/Fe] by $\sim$ 0.4 dex for stars with [Fe/H] > -0.5, and in [Al/Fe] by $\sim$ 0.45 dex for stars with [Fe/H] > -0.6. The increase in [Na/Fe] and [Al/Fe] at larger metallicities could be a sign of the contribution from AGB stars, similar to the [(C+N)/Fe] abundance pattern described in \S \ref{cno}. Sgr is also deficient in [P/Fe] by $\sim$ 0.2 dex for [Fe/H] > -0.6 and in [K/Fe] by $\sim$ 0.3 for [Fe/H] > -0.8.  As was the case for the $\alpha$ elements, Sgr stars with [Fe/H] < -0.8 merge with the high-latitude MW stars for [Al/Fe], [P/Fe], and [K/Fe].

The interpretation of the odd-Z element abundance patterns is similar to that of the $\alpha$-element abundance patterns described in \S \ref{alpha}. Sgr is more deficient in the mass-dependent [Na/Fe] and [Al/Fe] than in [P/Fe] and [K/Fe], which is a prediction of the top-light IMF scenario. The correlation of [Al/Fe] and [Fe/H] for Sgr stars with [Fe/H] > -0.5 suggests significant AGB contributions to the Al abundance of Sgr. This correlation is absent in [K/Fe], which is to be expected because K is produced in much lower quantities than Al in AGB stars (see yields from \citealt{Karakas2010}). 

\begin{figure*}[h]
\epsscale{1.}
\plotone{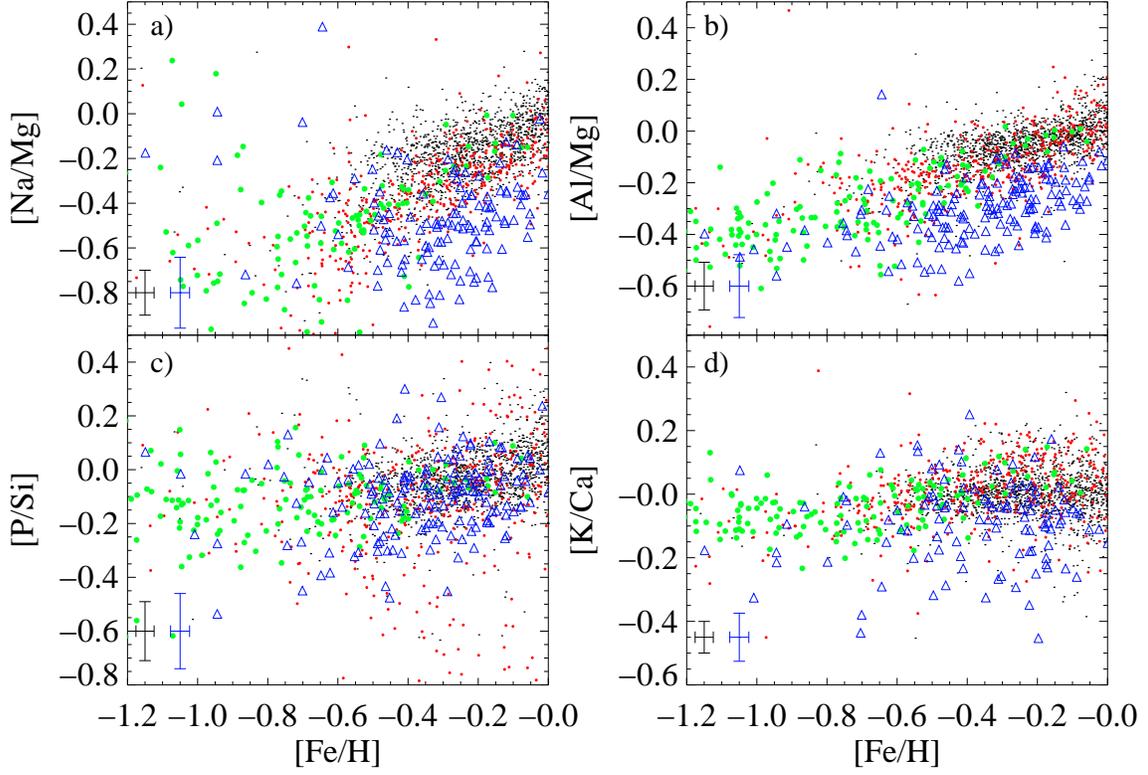} 
\caption{Plots of the odd-$Z$ element to $\alpha$ element abundance patterns for Sgr and the MW. Points are colored the same as in Figure \ref{cno_plot}.}
\label{fe_depend_plot}
\end{figure*}

The magnitudes of the deficiencies for the odd-Z elements are larger than the deficiencies in the $\alpha$-elements. Figure \ref{fe_depend_plot} shows the ratios of [Odd-Z/$\alpha$] for elements that are produced in similar burning processes. Sgr is deficient in [Na/Mg] and [Al/Mg] (carbon-burning products) by $\sim$ 0.25 and $\sim$ 0.35 dex, respectively. This suggests that the Type II SNe that enriched the gas from which the most recent generations of stars formed were more metal poor than the MW. Sgr is only slightly deficient (0.05-0.1 dex) in the [P/Si] and [K/Ca] ratios (oxygen-burning products). The yields of P and K are predicted to be less metallicity-dependent than Na and Al (see e.g., \citealt{Andrews2017} and the yields from \citealt{Chieffi2004}) so the abundance ratio for Sgr plotted in Figure \ref{fe_depend_plot} are consistent with more metal-poor Type II SNe enrichment.

\subsection{Fe-peak Elements}
\label{fepeak_elem}

The Fe-peak elements are the heaviest elements synthesized via nuclear fusion; the production of these elements occurs in explosive Si burning. This burning process occurs both in the cores of massive stars immediately preceding a Type II SNe explosion and in Type Ia SNe explosions. Type Ia SNe explosions release Fe-peak elements largely with the exclusion of the other lighter elements. This is understood from calculated yields (e.g., \citealt{Iwamoto1999}), as well as observations of the MW thick-thin disk transition in the [$\alpha$/Fe] vs. [Fe/H] abundance space (e.g., \citealt{Tinsley1979}). 

\begin{figure*}[h]
\epsscale{1.}
\plotone{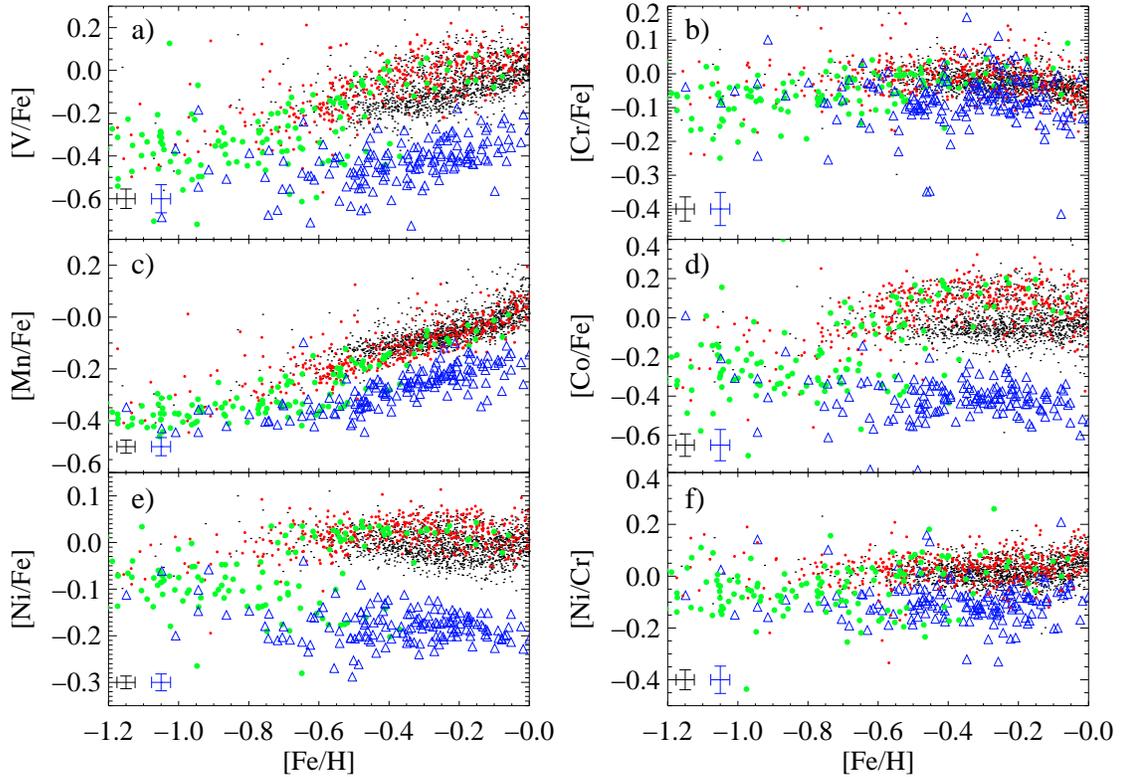} 
\caption{Chemical abundance patterns of the Fe-peak elements. Colors are the same as in Figure \ref{cno_plot}. }
\label{fe_peak_plot}
\end{figure*}

Figure \ref{fe_peak_plot} shows that Sgr is deficient in all of the Fe-peak elements, except for Cr, at metallicities above [Fe/H] = -0.8. The MW and Sgr share correlations of [V/Fe] and [Mn/Fe] with [Fe/H] although Sgr is deficient by $\sim$ 0.3 dex in both [V/Fe] and [Mn/Fe]. Mn is the only Fe-peak element to have been studied previously in Sgr. \citet{McWilliam2013} found that Sgr was likely deficient in [Mn/Fe], but was unable to draw a firm conclusion due to their small sample size and potential NLTE effects. We do not correct for any NLTE effects here. However, because we're comparing to APOGEE data for a MW sample of similar stellar parameters and yet there is no overlap between the Sgr and MW sample in Figure \ref{fe_peak_plot}, the differences in [Mn/Fe] cannot be due to departures from NLTE. 

Sgr is also deficient in [Ni/Fe] and [Co/Fe] by $\sim 0.2-0.3 $ dex. The separation of [Ni/Fe] between Sgr and the MW is large compared to the [Ni/Fe] scatter of the two samples, and remains so to metallicities down to [Fe/H] $\sim$ -0.9. [Ni/Fe] deficiencies have also been observed in Fornax and the LMC (\citealt{Lemasle2014} and \citealt{VanderSwaelmen2013}, respectively). Sgr is only slightly deficient in [Cr/Fe] ($\sim$ 0.05 dex). Sgr stars more metal poor than [Fe/H] = -0.8 once again merge with the high-latitude sample for all Fe-peak elements.

It is unlikely that a lack of Type Ia SNe contributed to the deficiencies of the Fe-peak elements in Sgr. If this were the case, then the [$\alpha$/Fe] abundance ratios should exhibit enhancements at [Fe/H] > -0.8 instead of overall deficiencies. The yields of all Fe-peak elements are dependent on metallicities to various degrees, including Fe (\citealt{Woosley&Weaver1995,Chieffi2004,Nomoto2006}). Elements such as Mn and V are more metallicity-dependent than Fe, as indicated by calculated yields as well as the MW abundance pattern. \citet{Andrews2017}, using the \citet{Chieffi2004} Type II SNe yields, showed that Co and Ni are also more metallicity-dependent than Fe, although the Ni yields do not reproduce observed chemical tracks in the MW. Cr is about as metallicity-dependent as Fe, so the [Cr/Fe] vs. [Fe/H] trends should look the same, regardless of the metallicity of the SNe. 

The [V/Fe]-[Fe/H] and [Mn/Fe]-[Fe/H] correlations of both the Sgr and MW samples can be accounted for by the metallicity-dependent yields of V and Mn in Type II SNe. As more metal-rich Type II SNe explode, the [Mn/Fe] and [V/Fe] abundance ratios become more enhanced. This may also imply that the fraction of Type II SNe ejecta to Type Ia SNe ejecta is similar for both Sgr and the MW, at least to the extent that Type Ia SNe cannot dominate. Yields of Type Ia SNe from \citet{Iwamoto1999} indicate that Mn is produced at around the solar [Mn/Fe] abundance. If Type Ia ejecta dominated much more than Type II SNe ejecta, the Sgr [Mn/Fe] abundance ratio would approach solar, instead of having roughly the same slope of the MW sample. The offset of [V/Fe] and [Mn/Fe] between Sgr and the MW can be explained by mass-dependent Type II SNe yields. Mn and V are produced in greater quantities relative to Fe in more massive Type II SNe (see e.g., \citealt{Chieffi2004} yields and \S \ref{flex_ce}). Therefore, [Mn/Fe] and [V/Fe] are expected to be lower in Sgr if it was devoid of massive stars polluting the gas that formed the most recent generations of Sgr stars. 

The [Co/Fe] and [Ni/Fe] abundance patterns cannot be easily understood with current yields. Chemical-evolution models using current Ni yields tend to produce tracks that rise with [Fe/H] and reach super-solar values (e.g., \citealt{Andrews2017}). We see no signs that the production of Ni is metallicity-dependent in both the MW and Sgr samples. Similarly, the yields of Co should be metallicity-dependent, but we do not see the correlations with [Fe/H] that are apparent in the [V/Fe] and [Mn/Fe] abundance patterns. However, Co is expected to be produced at sub-solar abundance ratios relative to Fe in Type Ia SNe. Therefore, low [Co/Fe] could be an indication that Type Ia SNe contributed more to the chemical enrichment of Sgr relative to Type II SNe than the MW. Whether this can happen in a way that still produces the same [Mn/Fe]-[Fe/H] correlation requires more precise Co yields. The production of Cr is not mass-dependent but is as metallicity-dependent as Fe in both Type II and Type Ia SNe. Therefore, the fact that Sgr is deficient in [Ni/Cr] is not surprising.

\subsection{Hydrostatic vs. Explosive Elements}
\label{hydro_explo}

One of the strongest arguments in favor of the steep or top-light IMF scenario from \citet{McWilliam2013} is that they found that Sgr appeared to be more deficient in the hydrostatic elements (O, Mg, and Al) than in the explosive elements (Ca, and Ti). The yields of the former are dependent on the mass of the progenitor whereas the explosive elements have little to no mass dependent yields. Deficiencies in the hydrostatic elements that are greater than the deficiencies in the explosive elements can occur if the gas that formed the Sgr stars more metal rich than [Fe/H] $\simeq$ -0.8 was lacking in ejecta from massive Type II SNe. 

\begin{figure*}[h]
\epsscale{1.}
\plotone{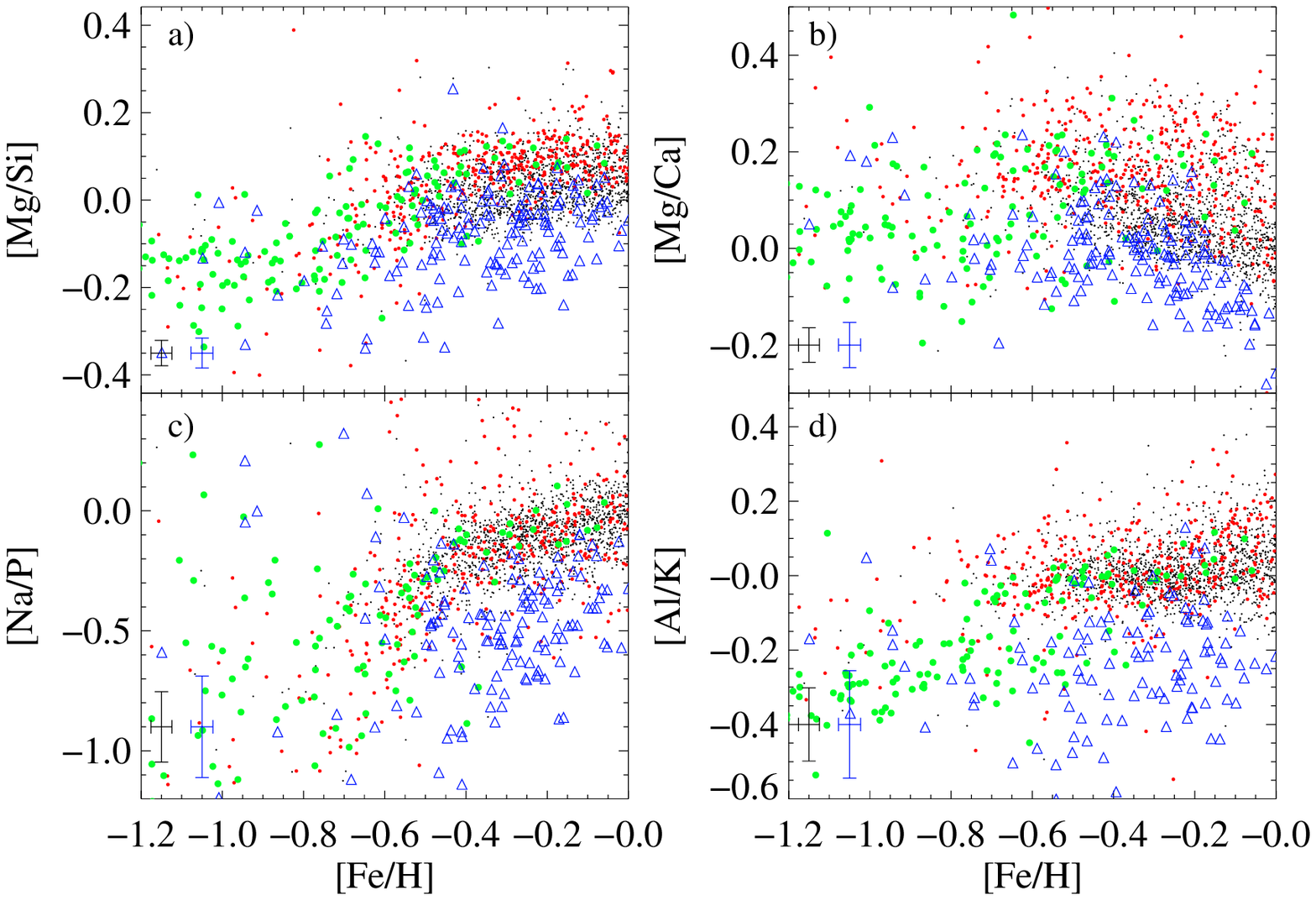} 
\caption{Elemental abundance ratios of hydrostatic to explosive elements. Colors are the same as in Figure \ref{cno_plot}. }
\label{hydro}
\end{figure*}

Figure \ref{hydro} shows the abundance ratios of various hydrostatic to explosive elements. Sgr is slightly deficient in both [Mg/Si] and [Mg/Ca], by $\sim$ 0.1 dex.  The yields of Mg, like O, depend more heavily on the mass of the Type II SN progenitor than Si and Ca. Therefore, deficient [Mg/Si] and [Mg/Ca] abundance ratios are expected in the top-light IMF scenario.

Ratios of the odd-$Z$ elements tell a similar story, although their interpretation is complicated by the fact that the yields of these elements are more strongly metallicity-dependent than the yields of the $\alpha$-elements. Yields of Na and Al are dependent on the mass of the progenitor, whereas yields of P and K are less so. Sgr is deficient in [Na/P] by $\sim$ 0.2 dex and Sgr by $\sim$ 0.2 dex in [Al/K] relative to the MW. Even when analyzing different hydrostatic-explosive element ratios, we reach the same conclusion as \citet{McWilliam2013}, that these deficient ratios suggest a top-light IMF scenario rather than SNe outflows. 

\subsection{Comparison to Mucciarelli et al. 2017}
\label{comp}

\begin{figure}[h]
\epsscale{1.}
\plotone{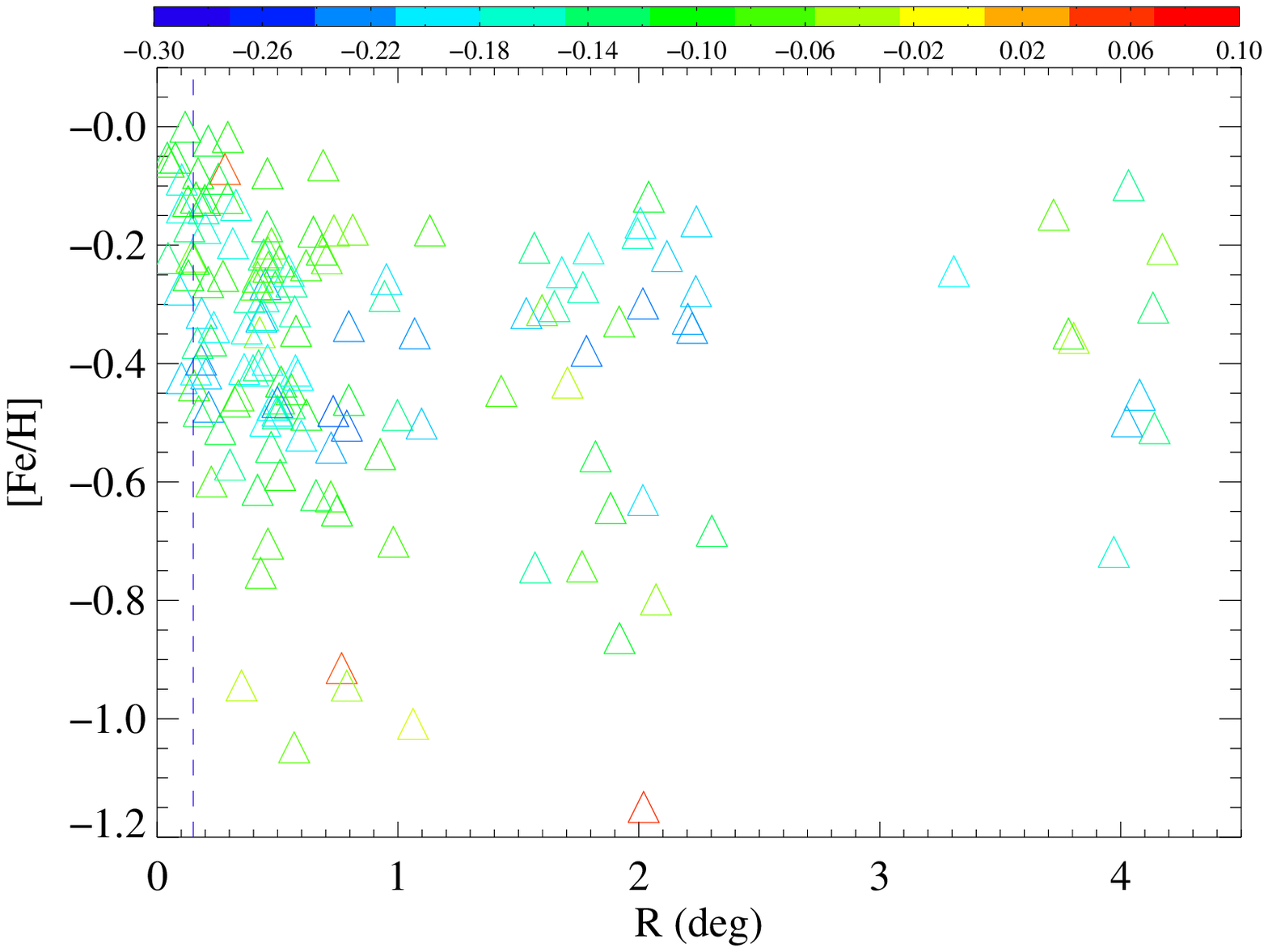} 
\caption{Metallicity as a function of projected distance from the center of Sgr. The points are colored by [$\alpha$/Fe]. The blue dashed-line indicates the extent of the region probed by \citet{Mucciarelli2017}. }
\label{grad}
\end{figure}

Recently, \citet{Mucciarelli2017} published Fe, Mg, Ca, and Ti abundances obtained from R $\sim$ 6600 Keck/DEIMOS spectra of 235 stars in the inner 9' of Sgr. They found similar $\alpha$-element deficiencies that we find in our sample for Sgr stars with [Fe/H] > -1.0. They also found a metallicity gradient across the inner 9', but claim that the $\alpha$-element trends are the same across the entire region. Figure \ref{grad} shows the metallicity of our Sgr sample plotted against projected radius. The points are colored according to their $\alpha$-element abundance. Our sample is more widely distributed across the dwarf than the sample of \citet{Mucciarelli2017}, and we only have 13 stars that are within 9'. While the study of gradients and MDFs of the APOGEE Sgr sample is beyond the scope of this study, we do see hints of a metallicity gradient extending to R $\sim$ 1 degree, (first explored in the APOGEE data in \citealt{Majewski2013}) and also find that the [$\alpha$/Fe] deficiencies for the metal-rich ([Fe/H] > -0.8) Sgr stars extend beyond R $\sim$ 4 degrees from the center of Sgr. The $\alpha$-element deficiencies do not appear to be unique to the center of Sgr.

\citet{Mucciarelli2017} were able to reproduce the observed $\alpha$-element deficiencies by using a chemical evolution model in which Sgr loses much of its gas to tidal stripping beginning at 7.5 Gyr ago, rather than a top-light IMF. While they are able to accurately match the [Mg/Fe] and [Ca/Fe] abundance patterns with their model, it is unclear whether such a model predicts the same relative deficiencies of the hydrostaic and explosive elements as shown in Figure \ref{hydro_explo}. In our simple chemical evolution model explained in \S \ref{flex_ce}, we do not take in to account stripping of gas, but note here that it is potentially an important factor in the chemical evolution history of Sgr. Further studies of the detailed chemical abundances of the Sgr streams will help to place more robust time constraints on the gas stripping of Sgr.

\subsection{Comparison to Other Dwarf Galaxies}

Similar deficiencies have been observed in Fornax \citep{Lemasle2014} and the LMC \citep{VanderSwaelmen2013}. \citet{Lemasle2014} found that Fornax was deficient in [$\alpha$/Fe] and enhanced in [Eu/Mg], just as \citet{McWilliam2013} found for Sgr. This led them to conclude that the more metal-rich Fornax stars formed from gas that was lacking in ejecta from massive Type II SNe. \citet{VanderSwaelmen2013} found the [Mg/Fe] and [O/Fe] are deficient relative to the MW in the LMC, but [Si/Fe], [Ca/Fe], and [Ti/Fe] are similar between the LMC and MW. From this they conclude that massive stars had a smaller contribution to the chemical enrichment of the LMC relative to Type Ia SNe and AGB stars.

\begin{figure*}[h]
\epsscale{1.}
\plotone{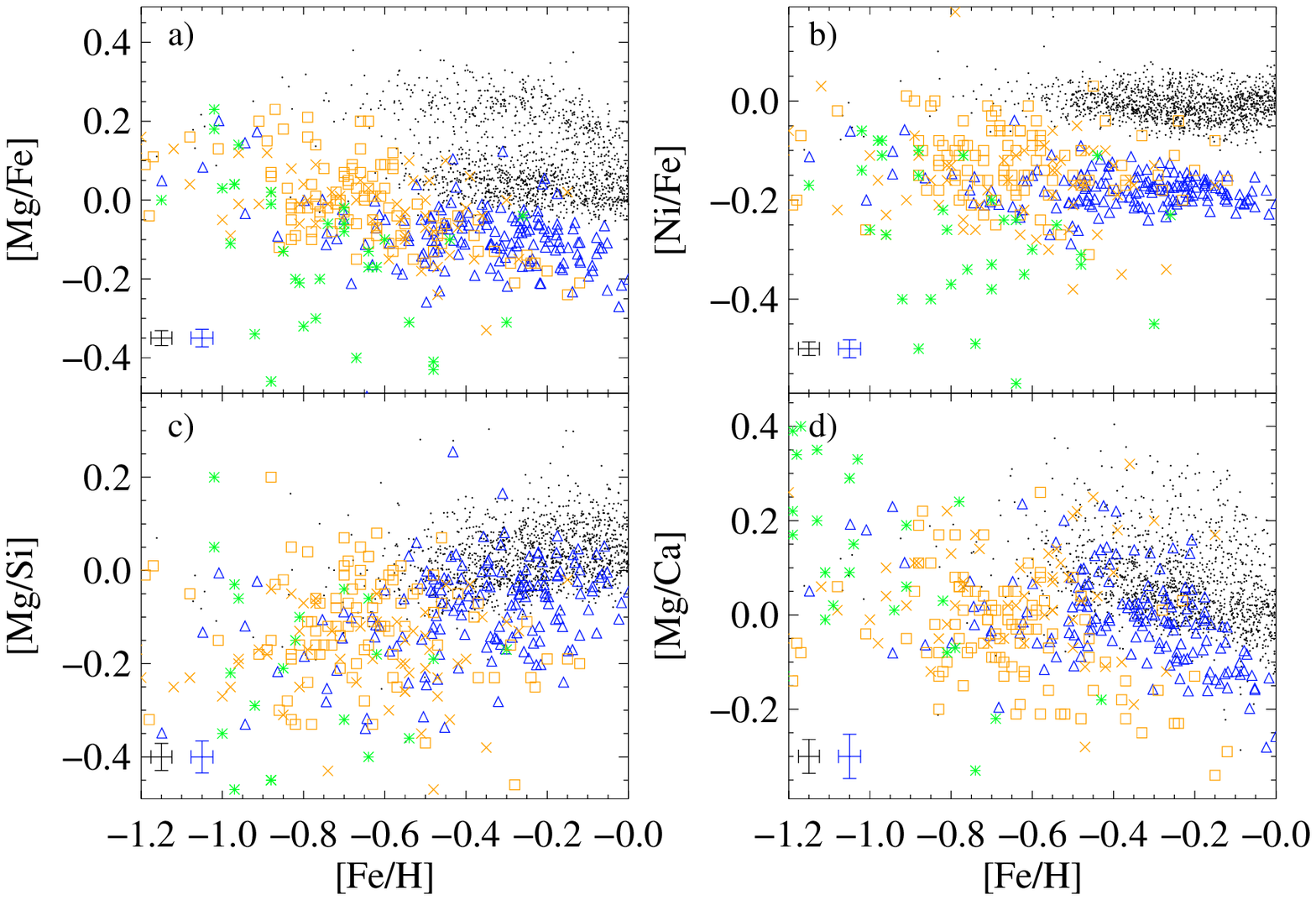} 
\caption{Plots comparing the Fornax sample from \citet{Lemasle2014} (green asterisks) along with the LMC disk (orange open boxes) and LMC bar (orange crosses) samples from \citet{VanderSwaelmen2013} to the MW disk and Sgr samples.}
\label{dwarf}
\end{figure*}

Figure \ref{dwarf} compares the Sgr sample, MW disk sample, LMC sample from \citet{VanderSwaelmen2013}, and Fornax sample from \citet{Lemasle2014}. It is clear from panel a and b that all of the dwarf galaxies are deficient in [Mg/Fe] and [Ni/Fe] relative to the MW for [Fe/H] > -0.8. The LMC shows deficient [Mg/Si] and [Mg/Ca] ratios, which we argue for Sgr implies the top-light IMF scenario. This comparison shows that Sgr is not unique among dwarf galaxies in its chemical abundance patterns and suggests that dwarf galaxies form stars according to a top-light IMF, possibly because they lack the more massive molecular clouds required to form massive stars (see e.g., \citealt{Oey2011}).

\subsection{Implications for Halo Accretion Scenarios}

When compared to the MW disk, bulge, and high-latitude stars, the chemical abundance patterns of Sgr are quite different. In most of the elemental plots shown thus far, almost no MW points coincide with the Sgr locus. Moreover, the APOGEE high-latitude sample, which is supposed to contain a large fraction of halo stars, has minimal overlap with Sgr in [Fe/H]. This implies that the MW halo was not formed via the accretion of Sgr-like dwarf galaxies as they are observed today. However, according to the Sgr SFH from \citet{Siegel2007} and the [C/N] abundances presented in Figure \ref{cno_plot_b}, the more metal-rich Sgr stars are only as old as $\sim$ 8 Gyr. If the accretion history of the MW was peaked at earlier times, as some work suggests (see e.g., \citealt{Robertson2005,Zolotov2010}), then metal-rich Sgr-like stars would not be found in the MW halo today as they hadn't formed yet.

As noted throughout the text, Sgr stars with [Fe/H] < -0.8 appear to join with the high-latitude sample. In many of the elements, the high-latitude+Sgr stars appear to form a chemical sequence. The high-latitude sample was designed to include predominantly halo stars. However, we're only looking at stars with [Fe/H] > -1.2. Therefore, we are comparing to the most metal-rich stars in the halo as the halo MDF is peaked at [Fe/H] $\simeq$ -2.1 and [Fe/H] $\simeq$ -1.5 (see e.g., \citealt{AllendePrieto2014,An2015,Fernandez-Alvar2017}). The high-latitude sample we compare to contains stars that are more deficient in [$\alpha$/Fe], [Al/Fe], and [(C+N)/Fe] relative to the thick disk and the canonical halo. Previous works suggests that these stars have chemical abundances and kinematics consistent with having been accreted and represent the "accreted halo" population (\citealt{Nissen2010,Nissen2014,Hawkins2015,Fernandez-Alvar2017}). A recent study of the APOGEE halo stars by \citet{Fernandez-Alvar2017} revealed that the outer halo is on average more $\alpha$-element poor than the inner halo. Because the more metal-poor Sgr stars join up with the chemical sequence of the high-latitude stars, we support the conclusion that the $\alpha$-poor, more metal-rich (-1.2 < [Fe/H] < -0.8) halo stars are indeed an accreted population. \citet{Kobayashi2014} suggested that this accreted population could have been formed via the top-light IMF scenario.

\section{Galactic Chemical Evolution with FlexCE}
\label{flex_ce}

We use a parameterized chemical evolution model (flexCE) developed by \citet{Andrews2017} to analyze how tweaking the star-formation history of a galaxy affects the detailed chemical-abundance patterns. FlexCE\footnote{FlexCE is publicly available from https://github.com/bretthandrews/flexCE} is a one-zone, open box chemical-evolution model that includes yields from Type II SNe (\citealt{Limongi2006} and \citealt{Chieffi2004}), Type Ia SNe (\citealt{Iwamoto1999}), and AGB stars (\citealt{Karakas2010}). \citet{Andrews2017} adopted a  ``fiducial'' simulation with parameters that produced results that best represent the MW [O/Fe] abundance distribution from \citet{Ramirez2013}.

We use flexCE to run five simulations to investigate what parameters may give rise to the Sgr abundance patterns:
\begin{itemize}
\item Fiducial simulation from \citet{Andrews2017}. This should match the MW abundance patterns.
\item Fiducial simulation with the upper mass-cutoff of a Kroupa IMF lowered from 100 $M_{\odot}$ to 30 $M_{\odot}$ (top-light IMF). 
\item Fiducial simulation with the star-formation efficiency factor reduced by an order of magnitude ($10^{-10} Gyr^{-1})$.
\item Fiducial simulation with the mass-loading factor of the outflow increased from 2.5 to 4.0 (galactic winds).
\item Fiducial simulation with both the decrease in star-formation efficiency and increased mass-loading factor.
\end{itemize}

\begin{figure*}[h]
\epsscale{1.}
\plotone{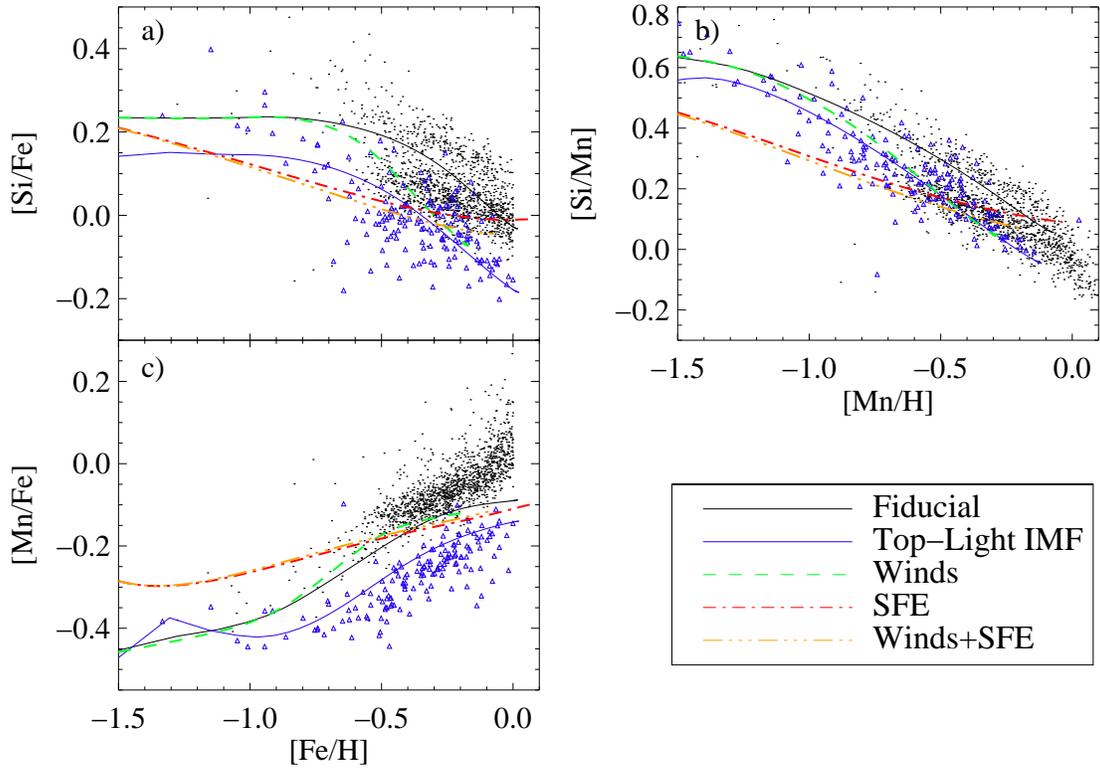} 
\caption{Select chemical abundance patterns with the chemical abundance tracks from flexCE. Coloring of the lines is indicated in the figure legend.}
\label{verne_disc}
\end{figure*}

We limit our analysis only to the elements for which the fiducial simulation matches the APOGEE MW abundances. The yields for many of these elements are uncertain, and flexCE (as well as other chemical-evolution models) do not properly predict the MW abundance patterns. Figure \ref{verne_disc} shows the results for Si and Mn. The expected chemical abundance tracks from each simulation are over-plotted. The solid black line is the fiducial simulation which should correspond to the APOGEE MW abundances for Si and Mn.

The purple solid lines in Figure \ref{verne_disc} represents the simulation with the upper mass-cutoff set to $30 M_{\odot}$. The green and red dashed lines show the tracks for the increased mass-loading parameter and reduced star-formation efficiency, respectively. The orange line shows the tracks for the reduced star-formation efficiency combined with the increased mass-loading parameter. The [Si/Fe] abundance patterns of Sgr appear to be well reproduced by both the upper mass-cutoff simulation and the higher mass-loading parameter simulation. However, the increased mass-loading does not properly reproduce the [Mn/Fe] abundance patterns in Sgr ([Mn/Fe] too high), whereas the upper mass-cutoff simulation does. The [Si/Mn] abundance pattern is well-reproduced by the upper mass cutoff simulation. The upper mass cutoff simulation reaches the metal-rich end of the Sgr sample in both [Fe/H] and [Mn/H], whereas the mass loading simulation falls short by $\sim$ 0.2 dex of enriching the metallicity to the observed values. The simulation with both the reduced star-formation efficiency and increased mass-loading exhibits more $\alpha$-element deficiency at the higher metallicity end than the reduced star-formation efficiency only, but also suffers from enriching the metallicity to the observed values of [Fe/H]. 

While this analysis is by no means a full, robust description of the chemical-abundance patterns of the Sgr system, it is suggestive that a top-light IMF formation scenario does produce chemical-abundance patterns corresponding to the Sgr tracks, at least for some of the elements we are able to study here. Not only are the tracks for Mn and Si reproduced, but the top-light IMF reaches metallicities as high as [Fe/H] = 0.0, as we see in the Sgr APOGEE sample. It is more difficult to get to [Fe/H] = 0.0 with strong outflows. \citet{Vincenzo2015} used a more detailed chemical evolution model for Sgr and found that they were able to better match the observed Sgr abundance patterns if they changed the IMF to reduce the amount of massive stars. As yield calculations improve, revised chemical evolution models applied to this APOGEE dataset will help to untangle the star formation history and chemical evolution of Sgr. 

A different theoretical approach is provided by cosmological models which include supernova feedback and  chemical evolution (e.g. \citealt{Zolotov2009,Font2006,McCarthy2012,Tissera2013,Tissera2014}) These models describe the assembly history of the stellar halos by the accretion of dwarf galaxies of different masses. Within this context, the early accretion of massive satellites could contribute with old and metal-rich stars \citep{Font2006} as well as with stellar populations with variety of the alpha-enhacements depending on the star formation histories of the accreted dwarf galaxies \citep{Tissera2014}. Hence, the more complex scenario for the halo formation could provide an explanation of the reported abundances for Sgr but this remains to be proven in detail.

\section{Conclusions}

We have provided detailed chemical abundances from APOGEE for 15 elements in 158 Sgr member stars. This is the largest, and most chemically extensive high-resolution survey of the Sgr galaxy to date. We find that Sgr is deficient, at various levels, in all the studied chemical-abundance ratios relative to Fe, which indicates that the most recent generations of Sgr stars with [Fe/H] $\gtrsim$ -0.8 formed from gas that was much less polluted with Type II SNe than the gas that formed stars in the MW disk and bulge. Evidence such as the deficient hydrostatic element to explosive element ratios ([Mg/Si], [Mg/Ca], [Na/P], and [Al/K]) suggests that the lack of Type II SNe ejecta stems from a top-light IMF rather than from outflows. This is similar to the conclusion reached by \citet{McWilliam2013}, who observed that Sgr is more deficient in O, Mg, and Al than it is in Ca and Ti.  Using the flexCE chemical evolution model from \citet{Andrews2017}, we have shown that the [Si/Fe] and [Mn/Fe] abundance patterns can be reproduced simply with an upper cutoff of the IMF, and that galactic winds cannot lower these abundance ratios and simultaneously enrich [Fe/H] to the observed high abundances in Sgr. Fornax and the LMC also show deficient hydrostatic/explosive element ratios and the top-light IMF scenario can also explain the abundance patterns of those dwarf galaxies.   

We also find that AGB stars are fractionally a much larger contributor to the chemical enrichment of Sgr than in the MW. We find clear signs of AGB enrichment beginning at [Fe/H] $\sim$ -0.6 in [(C+N)/Fe], [Na/Fe], and [Al/Fe], and that the abundance patterns of these elements approach the MW trend in the most metal-rich Sgr stars. This is consistent with \citet{McWilliam2013} who found strong [La/Fe] enhancements in Sgr beginning at [Fe/H] $\sim$ -0.6. A more robust analysis of $s$-process elements and carbon isotope ratios of Sgr will further constrain the AGB contributions to the chemical evolution history of Sgr.  

Sgr stars with [Fe/H] < -0.8 are chemically similar to the MW high-latitude sample. Work in the literature suggests that this high-latitude MW sample we compare to is predominantly made up of stars belonging to the "accreted" halo. These stars are less $\alpha$-element enhanced than the thick disk and canonical halo. The chemical similarity implies that the MW halo was formed, at least in part, through the accretion of Sgr-like dwarf galaxies at a time before chemical evolution could enrich the dwarf galaxies to higher values of [Fe/H] such as those observed in Sgr today.

\vspace{0.5cm}
\scriptsize{\emph{Acknowledgements.} We thank the anonymous referee for the useful comments that have improved this manuscript. Funding for the Sloan Digital Sky Survey IV has been provided by the
Alfred P. Sloan Foundation, the U.S. Department of Energy Office of
Science, and the Participating Institutions. SDSS acknowledges
support and resources from the Center for High-Performance Computing at
the University of Utah. The SDSS web site is www.sdss.org.

SDSS is managed by the Astrophysical Research Consortium for the Participating Institutions of the SDSS Collaboration including the Brazilian Participation Group, the Carnegie Institution for Science, Carnegie Mellon University, the Chilean Participation Group, the French Participation Group, Harvard-Smithsonian Center for Astrophysics, Instituto de Astrof{\'{\i}}sica de Canarias, The Johns Hopkins University, Kavli Institute for the Physics and Mathematics of the Universe (IPMU) / University of Tokyo, Lawrence Berkeley National Laboratory, Leibniz Institut f{\"u}r Astrophysik Potsdam (AIP), Max-Planck-Institut f{\"u}r Astronomie (MPIA Heidelberg), Max-Planck-Institut f{\"u}r Astrophysik (MPA Garching), Max-Planck-Institut f{\"u}r Extraterrestrische Physik (MPE), National Astronomical Observatory of China, New Mexico State University, New York University, University of Notre Dame, Observat{\'o}rio Nacional / MCTI, The Ohio State University, Pennsylvania State University, Shanghai Astronomical Observatory, United Kingdom Participation Group, Universidad Nacional Aut{\'o}noma de M{\'e}xico, University of Arizona, University of Colorado Boulder, University of Oxford, University of Portsmouth, University of Utah, University of Virginia, University of Washington, University of Wisconsin, Vanderbilt University, and Yale University.

D.A.G.H. was funded by the Ram\'on y Cajal fellowship number RYC$-$2013$-$14182. D.A.G.H. acknowledge  support  provided  by  the  Spanish  Ministry  of  Economy  and  Competitiveness (MINECO) under grant AYA$-$2014$-$58082$-$P. T.C.B. acknowledges partial support for this work from the National Science Foundation under Grant No. PHY-1430152 (JINA Center for the Evolution of the Elements). D.M. is supported by the Basal CATA through grant PFB-06, and the Ministry for the Economy, Development, and Tourism, Program ICM through grant IC120009, awarded to the MAS, and by FONDECYT grant No. 1130196. SV gratefully acknowledges the support provided by Fondecyt reg. n. 1170518. S.R.M. acknowledges NSF awards AST-1109718, AST-1312863, and AST-1616636.  C.A.P. is grateful for support from MINECO for this research through grant AYA2014-56359-P. JF-T., B.T., D.G. and S.V. gratefully acknowledge support from the Chilean BASAL  Centro de Excelencia en Astrof\'isicay Tecnolog\'ias Afines (CATA) grant PFB-06/2007.

\normalsize

\bibliographystyle{apj}
\bibliography{ref_og.bib}
 
\end{document}